\documentclass[aps]{article}
\usepackage{geometry}                
\usepackage{graphicx}

\usepackage{amssymb}
\usepackage{epstopdf}
\usepackage{amsmath}

\usepackage{amsfonts}
\usepackage{bm}
\usepackage{hyperref}

\DeclareMathAlphabet{\mathpzc}{OT1}{pzc}{m}{it}

\title{Two Experimental Tests to Distinguish Decoherence from the Slicing Theory of Measurement}
\author{Clifford Chafin\\\ \small{Department of Physics, North Carolina State University, Raleigh, NC 27695} \thanks{cechafin@ncsu.edu}}

\begin{document}
\maketitle
\begin{abstract}
Here we propose a pair of experiments to distinguish the recently proposed ``slicing theory'' of quantum measurement, which gives a transient many worlds picture, and decoherence.  Since these two theories are essentially ``opposites'' in their approach and both claim to arise from the many body Schr\"{o}dinger equation itself, there is no chance of them being equivalent representations of the same reality.  It will be explicitly shown that each theory gives very different answers to the questions of back reaction and revival of phase effects after measurement. We suggest that the kinds of isolated systems now possible in optical traps is now sufficient to generate a selective distinction between these two theories.  In particular we show that the slicing theory gives examples of interference from ``revival of histories'' in controlled examples but no back reaction on the measurement devices whereas decoherence gives the opposite.  
\end{abstract}

\section{Introduction}
Since the early days of the Schr\"{o}dinger equation \cite{Schrodinger} there was some hope that eventually all the ad hoc probabilistic formalism and axiomatic approaches would go away and we would be left with an initial value problem akin to a classical field theory \cite{Jammer}.  Axiomatic approaches like those of von Neumann \cite{vN} are appealing to mathematicians but physicists are generally looking for a continuous mapping that holds for all times in a fashion that includes the observers not depends on them for results.  Decoherence \cite{Schlosshauer} has evolved out of these ideas to suggest that if we could describe the many body state adequately, such a resolution would occur and the apparently probabilistic measurements are the ``dephasing'' of the system into particular measurement choices.  This is the essence of the ``einselection'' process.  Numerical approaches have given some mixed results and sometimes require some non unitary evolution aspects to achieve measurement or some analog of thermalization.  Conservation laws are then presumably buried in the back reaction of the medium.  It should be noted that such a process must occur extraordinarily rapidly and causality issues arise with larger measurement devices.  

Another idea was the ``many worlds'' theory \cite{Everett} which is almost the opposite of the decoherence picture.  Instead of a single observable reality, at each ``measurement'' (defined by observers or some less well specified action of the quantum system with its environment) causes a branching of the reality of the system into many copies of the original space with completed measurements that show up in the probabilities given by the Born interpretation.  Objections to this theory is that is seems to create a bifurcation of worlds that are unobservable thus unscientific.  Other proposals included Bohm's pilot wave theory \cite{Bohm} which was a kind of predecessor to decoherence.  The quantum aspect of the system and the reality of it, in the form of point particles existing together with the wave guiding the particle into channels.  The hope was always that this would be deterministic once some small enough scale of data was evident.  

The recent slicing theory is built on the observation that the classical world we live in 1. is built from objects far more specialized than any quantum eigenfunction and 2. the ``universes'' they describe occupy tiny fractions of the many body quantum space available to the system.  When delocalized one particle wavefunctions interact with such a world we can see that there is a many worlds type bifurcation at work \cite{Chafin-pip}.  This situation is, however, transient and dependent on the presence of condensed matter.  Gases by themselves become truly delocalized quantum objects on short time scales since the particular localizations necessary to build up a world that appears as a 3D classical one, from a Fock space tower of 3N-D wavefunctions, require large net mass to retain this property for long times.  

It should be noted that both slicing and decoherence make claims as to what the true many body Schr\"{o}dinger evolution is and therefore there is an objective difference in these two points of view.  The purpose of this article is to show that there are experimental procedures to distinguish them.  This author claims that logical investigations alone are sufficient to show that decoherence should not be a valid contender but nothing is more compelling than data.  We will now describe the proposed experiment.

\section{Discussion}
In the slicing theory a particle is adsorbed to a solid surface generally with the release of one or more photons.  When the incident particle is delocalized it can adsorb at many sites at many times.  The coordinate of the incident particle wavefunction $\psi(x)$ and any released photon fields $A(y)$ then provide coordinates for the macroscopic body to ``slice'' into long lasting independent histories that do not bleed into each other for long times (due to the relatively large mass of the bodies).  The resulting wavefunctions (neglecting photon coordinates) of the resulting body is (at the moment of adsorption) $\sum_{i}\Psi(X)b(x-x_{i},t_{0})$ where we have assumed that all incident particles can be thought of as having been adsorbed at time $t_{0}$.  The sum is over the set of discrete sites on the surface $x_{i}$ and the ``bump functions'' $b(x)$ are localized clusters of the center of mass of the adsorbed atom which take the value $\psi(x)$ at the moment of adsorption.  Electronic coordinates have been suppressed in this picture so that the $X$ coordinates are the center of mass coordinates of all the other atoms in the solid.  The large mass of the body and its localization of its net center of mass and many angular motions cause each copy to now evolve with slightly increased mass and shifted moment of inertia as a set of classical bodies.  Each body can be indexed by the location of its adsorbed particle and denoted as $\mathcal{O}_{i}$ now has a relative phase assignment $\psi(x_{i})$.  

The interesting question is what happens if we provide a photon flux to all these spaces so that they all release the adsorbed particle.  Note that shining a photon on the system does not automatically do this.  Some of it will almost certainly miss the body or fail to be absorbed.  Furthermore, if we tailor the choice of shining the light on the system to some location on the screen then the light is correlated with it and some of the sliced copies $\mathcal{O}_{i'}$ will not experience any light.  We must shine the light on the system with no such choice dependent observations.  However, it is not clear that we cannot observe the system before we do this.  The role of ``observation'' is fundamental in traditional quantum mechanics but has been hard to incorporate into the theory in an intrinsic manner.  The slicing theory adopts the point of view that the kinds of condensed matter we are confronted with, for whatever reason, is very specialized and not well described at low T by the ground state of the Hamiltonian.  In this case we are interested in when the interaction of such macroscopic condensed matter is made to be correlated so that past history is now partitioned from further superpositions and interaction.  

As an example, consider a sufficiently isolated screen that has so partitioned itself into histories across the coordinate label of an incident particle.  Now let us consider the most gentle interaction of the system with the outside world through a single photon that reflects off of it and is absorbed by external solid matter.  For a first scenario, let it be focused through a lens to a different point on the surface or the surface be made of chemically changing material like film.  In this case there is clearly a correlation of the screen with its collapsed site to the outside solid object.  One can however consider the case of unfocused light on a more generic surface.  In this case, it is not clear if the photon is absorbed in a local or delocalize fashion and if there is any remnant of local activity corresponding to it that could indicate the location of the collapse site on the screen.  Certainly, there are partially absorbing materials and finite ones where part of the incident flux is lost.  Since the foundation of slicing theory is that the persistent localization of atomic cores is the foundation of classical matter and quantum measurement, it is possible that the localization of a photon's effect that does not endure in the atomic positions will not be sufficient to produce durable slicing of the objects into independent histories.  The delocalization may diffuse into the kinds of many body motion best understood as ``heat.''
These considerations suggests some aspects of slicing must remain a topic for future investigation. 
\begin{figure}
\begin{centering}
   \includegraphics[width=2in]{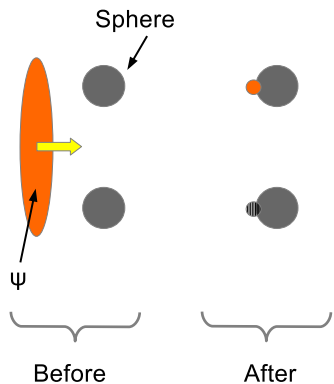}%
 \caption{\label{packet} A wavefunction packet incident on two small isolated spheres and, the aftermath, a particle adsorbed to the sphere with a corresponding site on the other sphere that is adsorbed in a parallel slice. }
 \end{centering}
 \end{figure} 

\section{Proposed Experiments}
The device we propose here is to be held to a high standard of isolation in that it should be in a cold container with highly reflective walls and high vacuum.  Instead of a screen let the solid objects be two solid balls of exactly equal net atomic number $N$ that are levitated optically or magnetically in the chamber as in fig.\ \ref{packet}.  We now send a delocalized particle of mass $m$ so that is incident on both of these balls at time $t_{0}$.  The ejected photon(s) then are absorbed (in a delocalized fashion) by the walls or are allowed to leave the cavity and continue indefinitely in infinite empty space.  Next, at time $t_{1}$, we send in a lone photon with energy sufficient to eject the particle.  This should be sufficiently localized in time so that the process of ejection at the two balls must be approximately synchronous.  The phases of the particle in each slice evolves much more rapidly while  being tied to the mass of the spheres.  If the mass of the spheres are exactly the same and cold enough that the thermal mass correction, $\Delta kTN$ between the two spheres, is small enough then the initial phase difference at the moment of ejection is going to be the same as during the particle capture, $\Delta\theta_{t_{0}}=\Delta\theta_{t_{1}}$.  The ejection rate of the particle can be tuned to have a low velocity and narrow spread if the photon that releases it in both slices is just above threshold.  

To determine the amplitude of the resulting flux we need to know how much of the photon strikes the spheres and how much is absorbed by the binding site in each slice.  The efficiency of this process will decide how often the particle in both slices is simultaneously ejected.  If this efficiency of ejecting from one of these locations is $\epsilon$ then the probability of a flux from both slices that can then interfere and effectively ``merge'' the two slices will be proportional to $\epsilon^{2}$.  This is an example of ``inverse collapse'' where some of the branching done and indexed by the incident particles coordinates are now rejoined.  We can now look for traces of an interference pattern on some external screen.  On top of this will be the $\epsilon$ scale larger non-oscillating distribution made from single ejection events.  The resulting interference pattern will be a function of the relative phase of the ejected sources.  For reasons stated, this may be a delicate effect that is hard to reproduce over many events which is what we need to get sufficient data to be certain that this effect is occurring.  The variation in the energy between the spheres $\Delta E\sim\Delta kTN$ gives a variation in the rate of phase advance as these move forwards in time.  For this not to disrupt the phase difference we need the difference in time to be smaller than a frequency oscillation induced by it, $t_{1}-t_{0}\ll\hbar/\Delta E$.  

Now let us compare this result with that of what we would expect from decoherence.  This presumes there is only one solution so no interference between different histories are there to be detected.  The measurement process is assumed to be ``thermodynamic'' in that it assumes that revival of past information is essentially impossible.  Aside from a null result in this revival experiment we can also investigate the roll of back reaction.  We know that the usual spacetime induced symmetries enforce conservation laws in many body quantum mechanics.  This is not as immediately obvious as in the classical case because the space is 3N dimensional but when we define translation, rotation, etc.\ as the 3D actions simultaneously acting over each 3D subspace the results hold.  In the case of a particle ``collapsing'' on a screen, this general results in an apparent violation of conservation of center of mass, momentum and angular momentum.  Over many measurements this averaged out but for decoherence this is a problem.  The resolution must be in the back reaction of the material to the shift induced in the position and momentum of the incident wavefunction.  In the slicing case, no such problem arises since these quantities are locally absorbed the the material or shared among the many sliced device paths $\mathcal{O}_{i}$.  
\begin{figure}
\begin{centering}
   \includegraphics[width=2in]{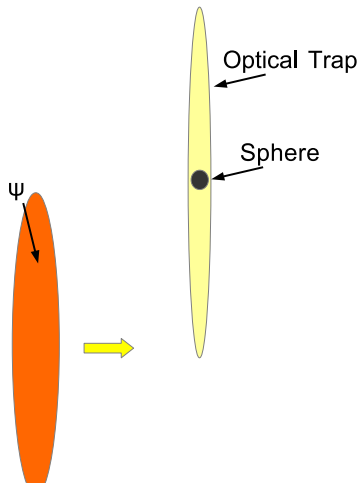}%
 \caption{\label{longtrap} A wavefunction packet incident on a long quasi-1D optical or magnetic trap with a sphere suspended in it.}
 \end{centering}
 \end{figure} 

In the decoherence case, such a reaction has causality problems and a bigger problem of being unspecified in how it occurs so that knowing what to look for as fluxes of momentum traverse the medium on the way to equilibration is unclear.  When the particle mass $m$ is not overwhelmingly smaller than the absorbing solid we have a chance of observing such a reaction.  The momentum recoil is proportional to $m/M_{sphere}$ but, for center of mass shifts, the displacement can be very large.  Consider a long, $2L$, 1D optical trap with very little lateral forces on the spheres as in fig.\ \ref{longtrap}.  Place a measurement sphere in the center of the trap and send in a wavefunction packet $\psi(x)$ that has width $2L$ with nearly uniform density and velocity field and with the center of the mass directed at one end of the trap.  When the particle is measured by the sphere, the shift in position must be to a location $\Delta x=\frac{m}{M}L$ from the center.  When $L$ is sufficiently long this can produce a measurable shift.  For an optical resolution of the system that is on the order of the sphere size $d$ and when $d\approx\Delta x$ the result is observable.  This provides a second test of decoherence versus slicing as an ``emergent'' theory of measurement from the Schr\"{o}dinger dynamics itself.  











\section{Conclusions}
The precise formulation of quantum reality as relates to measurement has often been viewed as a philosophical add-on to an otherwise completely sufficient positivist procedure.  Further digging has been suggested to be even meta-physical.  This point of view has neglected the extent to which formalism is often more of a set of traditions, with many implicit assumptions, than a true axiomatic endeavor.  Many of the areas in physics where we are stalled in progress ranging from quantum nonequilibrium dynamics to a unification of quantum mechanics with gravity are so because we lack a sufficiently clear picture of the underlying reality.  By producing a unification of classical and quantum dynamics that gives measurement as part of the same evolution, we will likely have new inspirations on these problems just as Newton's unification of the dynamics of heaven and Earth did around 350 years ago. 

Furthermore, this article should suggest that we are now at a threshold, already believed to be true to many, where a finer understanding of the measurement process is necessary.  The author of the original slicing theory is the author of this article so is clearly biased towards it over the decoherence point of view.  To be fair, he labored long attempting to make some analog of decoherence work in a consistent fashion with the various physical scenarios in which he wanted to make progress and finally concluded that it was hopeless and another approach was necessary.  Results from this are already underway in the areas of kinetics and equilibration.  Given the golden age of controlled measurement in which we live, perhaps some clever experimentalist will resolve this conundrum and open the door to a new era of progress on long frustrating problems.


\begin{thebibliography}{99}\footnotesize
\bibitem{Bell} J. S. Bell. On the problem of hidden variables in quantum mechanics, Rev. Mod. Phys., 38,  (1966).

\bibitem{Bohm}  D. Bohm, A suggested Interpretation of the Quantum Theory in Terms of Hidden Variables, I. Physical Review 85 (2): 166Ð179, (1952).

\bibitem{Chafin-pip}
C. Chafin. The Slicing Theory of Quantum Measurement: Derivation of Transient Many Worlds Behavior, Prog. in Phys.,  (in press)
(2015), quant-phys/arxiv.org/abs/1410.8238.

\bibitem{Everett} H. Everett., `Relative state' formulation of quantum mechanics. Rev. Mod. Phys., v.\,29, (1957).

\bibitem{Jammer} Max Jammer. \emph{The Philosophy of Quantum Mechanics; the Interpretations of Quantum Mechanics in Historical Perspective}. Wiley, (1974).

\bibitem{Schrodinger}
E. Schr\"{o}dinger. Die gegenw\"{a}rtige Situation in der Quantenmechanik. Naturwissenschaften v.\,23 (49), (1935).

\bibitem{Schlosshauer} M. Schlosshauer. Decoherence, the measurement problem, and interpretations of quantum  mechanics.  Rev. Mod. Phys., v.\,76, October (2004).
 
\bibitem{vN} J. von Neumann. \emph{Mathematical Foundations of Quantum Mechanics}, Princeton University Press, Princeton, (1955).

\end{thebibliography}
\end{document}